\documentclass{PoS}

\newcommand{\Kzs}{\mbox{$\mathrm {K^0_S}$}}
\newcommand {\mim} {\mbox{$ \mu {\rm m}$}}
\newcommand{\Jpsi} {\mbox{J\kern-0.05em /\kern-0.05em$\psi$}}

\title{ALICE status and highlights}

\ShortTitle{ALICE status and highlights}

\author{{Juergen SCHUKRAFT}  \\
      For the ALICE Collaboration\\
        CERN, Geneva\\
        E-mail: \email{jurgen.schukraft@cern.ch}}

\abstract{
 After close to 20 years of preparation, the dedicated heavy ion experiment ALICE took first data with proton collisions at the LHC starting in November 2009. This article summarizes initial operation and performance of ALICE at the LHC as well as first results from collisions at 900 GeV and 7 TeV.}

\FullConference{35th International Conference of High Energy Physics\\
		 July 22-28, 2010\\
		 Paris, France}

\begin{document}

\section{Introduction}
ALICE, which stands for A Large Ion Collider Experiment, is very different in both design and purpose from the other experiments at the LHC. Its main aim is the study of matter under extreme conditions of temperature and pressure, i.e. the Quark-Gluon Plasma, in collisions between heavy nuclei. With an energy up to 30 times higher than RHIC, the current energy frontier machine for heavy ion collisions at BNL, we expect both a very different type of QGP, e.g. in terms of initial temperature, lifetime and system volume, and an abundance of hard signals like jets and heavy quarks which serve as probes to study QGP properties. Data taking with pp (and later p-nucleus) is important for ALICE primarily to collect comparison data for the heavy ion program. Therefore our primary goal in 2010 was to collect about $10^{9}$  minimum bias (MB) collisions under clean experimental conditions (low event pile-up), which will provide sufficient comparison statistics for the first heavy ion run expected later this year. However, given the specific and complementary capabilities of ALICE, a number of measurements concerning soft and semi-hard QCD processes are of interest in their own in pp collisions and are part of the initial physics program~\cite{Carminati:2004fp,Alessandro:2006yt}. In addition, this large MB sample will provide a detailed characterisation of global event properties over a range of LHC energies, which will be very useful for tuning Monte Carlo generators to better describe the QCD background underlying searches for new physics.

\section{Detector performance and operation}
ALICE consists of a central part, which measures hadrons, electrons and photons, and a forward spectrometer to measure muons. The central part, which covers polar angles from $45^0$ to $135^0$ over the full azimuth, is embedded in the large L3 solenoidal magnet. It consists of an inner tracking system (ITS) of high-resolution silicon detectors, a cylindrical TPC, three particle identification arrays of Time-of-Flight (TOF), Cerenkov (HMPID) and Transition Radiation (TRD) counters, and two single-arm electromagnetic calorimeters (high resolution PHOS and large acceptance EMCAL). The forward muon arm ($2^0-9^0$) consists of a complex arrangement of absorbers, a large dipole magnet, and 14 stations of tracking and triggering chambers. Several smaller detectors for triggering and multiplicity measurements (ZDC, PMD, FMD, T0, V0) are located at small angles. 
The main design features include a robust and redundant tracking over a limited region of pseudorapidity, designed to cope with the very high particle density of nuclear collisions, a minimum of material in the sensitive tracking volume (10\% radiation length between vertex and outer radius of the TPC) to reduce multiple scattering, and several detector systems dedicated to particle identification over a large range in momentum~\cite{JSPLHC2010,Evans:2009zz}.

The layout of the ALICE detector and its eighteen different subsystems are described in detail in~\cite{ALICEdet}. The experiment is essentially fully installed, commissioned and operational, with the exception of the two systems (TRD and EMCAL) which were added more recently and are only now nearing the end of construction. Both systems have currently about 40\% of their active area installed and will be completed during the winter shutdown in 2011 (EMCAL) and 2012 (TRD). 

The very first pp collisions where observed in ALICE on November 23 2009.
The many years of preparation, analysis tuning with simulations, and detector commissioning with cosmics during much of 2008/9 paid of with most of the detector components working with collisions 'right out of the box' and rather close to performance specifications. Within days all experiments could show first qualitative results and the first phase of LHC physics, often referred to as the 'rediscovery of the standard model', was getting under way. The various members of the 'particle zoo' created in pp collisions made their appearance in ALICE in rapid succession, from the easy ones 
($\pi, K, p, \Lambda, \Xi, \phi,..$) in 2009 to the more elusive ones when larger data sets were accumulated early 2010 ($K^*, \Sigma^*, \Omega$, charmed mesons, $\Jpsi$, ..). 

\begin{figure}[!t]
%\centerline{\includegraphics[width=0.95\textwidth]{TPCdedx.eps}}
\centerline{\includegraphics[width=0.95\textwidth]{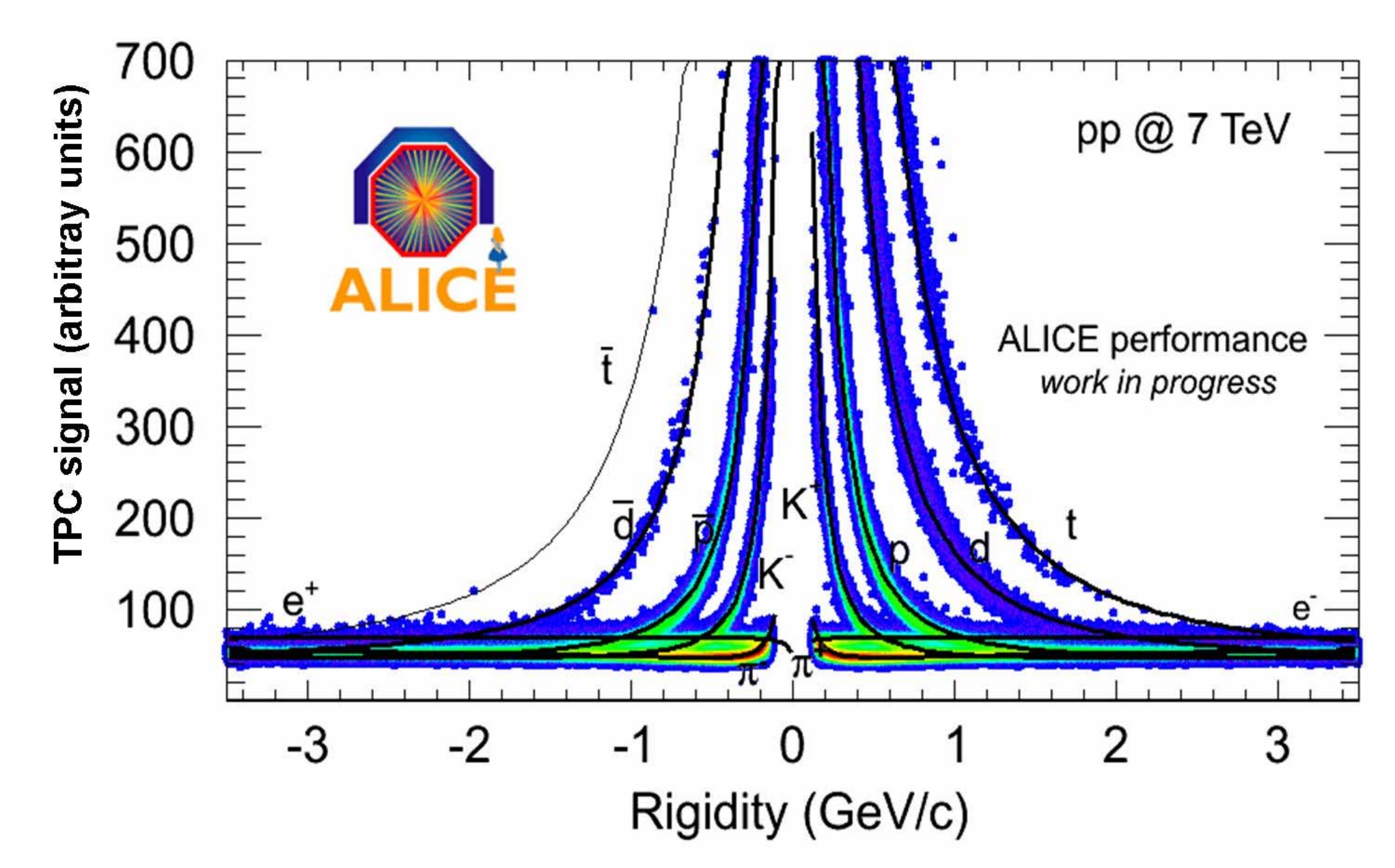}}

\caption{ Energy loss distribution versus rigidity for primary and secondary particles reaching the TPC. The lines overlaid on the distribution correspond to the expected energy loss for different particle species.}
\label{TPCdedx}
\end{figure}

As an example of the initial performance, the energy loss distribution in the TPC~\cite{Alme:2010ke} is shown in Figure~\ref{TPCdedx} versus rigidity, separately for positive and negative charges, demonstrating the clear separation between particle species reached in the non-relativistic momentum region. After careful calibration with radioactive Krypton injected into the gas volume and with cosmics, the energy loss resolution is about 5\%-6\%, i.e. at design value. Note that in this plot tracks are not required to point precisely back towards the vertex and therefore many secondaries produced in the detector material are included, which explains the different rates for positive and negative protons and light nuclei.

At the time of this conference, ALICE has collected more than 400 M  MB triggers and some 15 M muon triggers at 7 TeV, corresponding to a total integrated luminosity of a few \mbox{${\rm nb}^{-1}$}. At 900 GeV, some 10 M (300 k) MB triggers  were accumulated in 2010 and 2009, respectively. Since a few weeks the experiment is running in a special low luminosity mode, in which the LHC beams are separated by $ 3-5 \sigma$ in our interaction region, to keep event pile-up below a few percent per bunch crossing. This mode of operation turned out to be extremely stable, with a vertex distribution which is identical to the one for centred beams and, so far at least, without any observable impact on beam dynamics.

\section{Physics Results}

\begin{figure} \centerline{\includegraphics[width=0.95\textwidth]{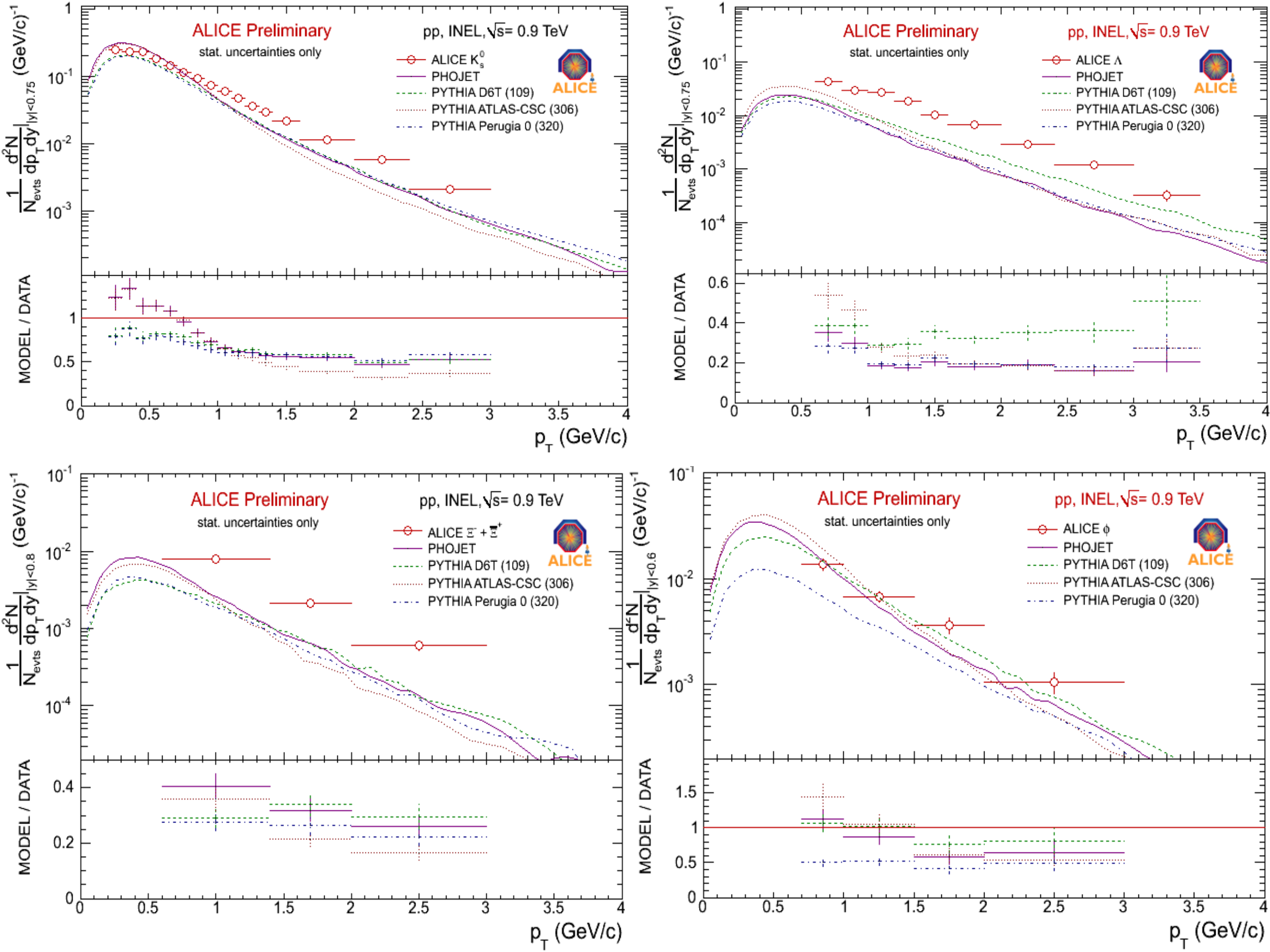}}

\caption{Yield of strange particles versus transvese momentum and comparison
with Phojet and different Phythia tunes (clockwise from top left:$\Kzs, \Lambda, \phi, \Xi^+ + \Xi^-$. The panel below each spectrum shows the ratio between model (Phojet and various Phytia tunes as indicated in the figures) and data.}
\label{f:strange}
\end{figure}

\begin{figure} \centerline{\includegraphics[width=0.95\textwidth]{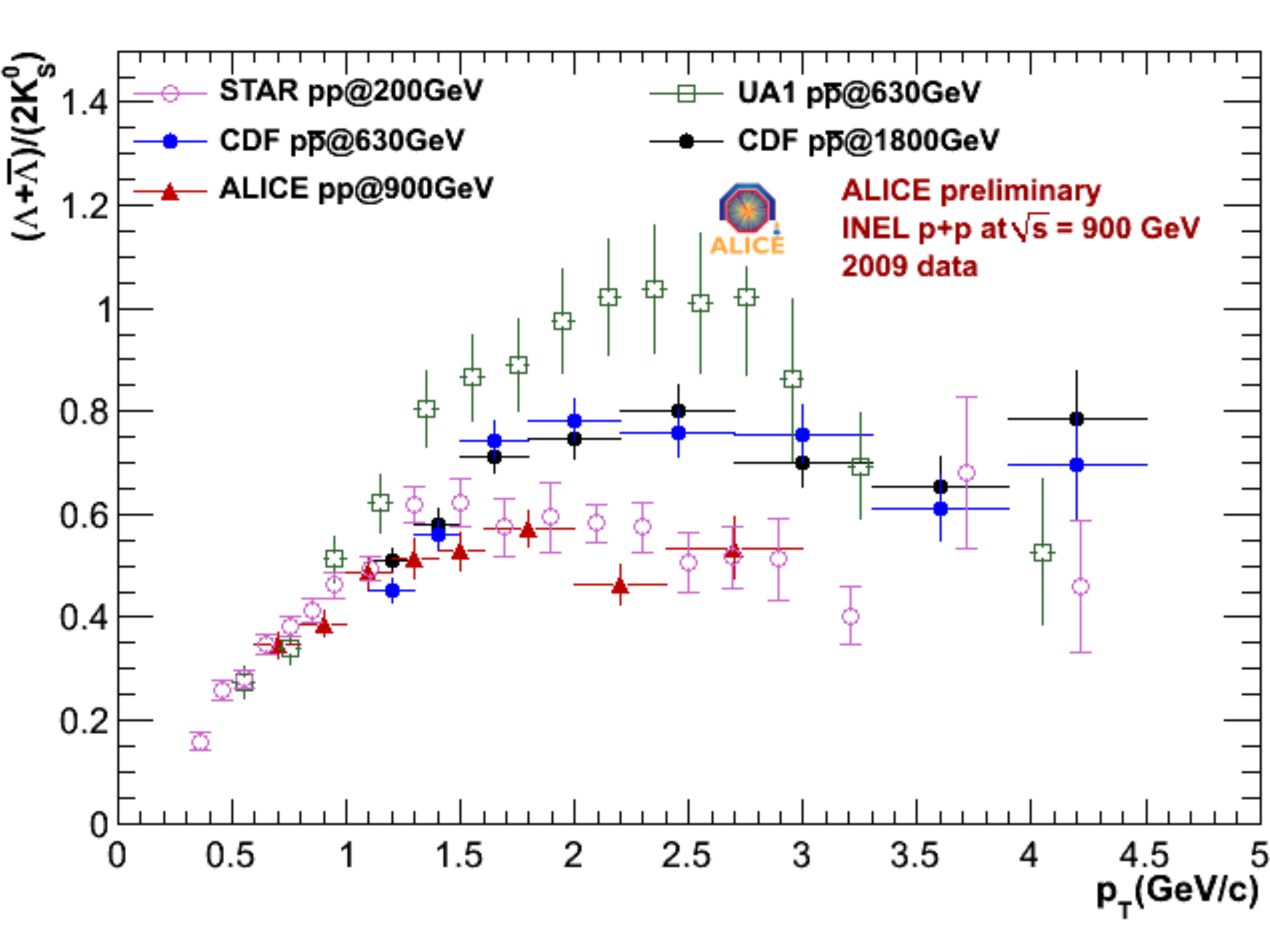}}

\caption{$\Lambda$ to $\Kzs$ ratio as function of transverse momentum for ALICE data at 900 GeV compared to data from the STAR, UA1 and CDF collaborations.}
\label{f:klambda}
\end{figure}

\begin{figure} \centerline{\includegraphics[width=0.95\textwidth]{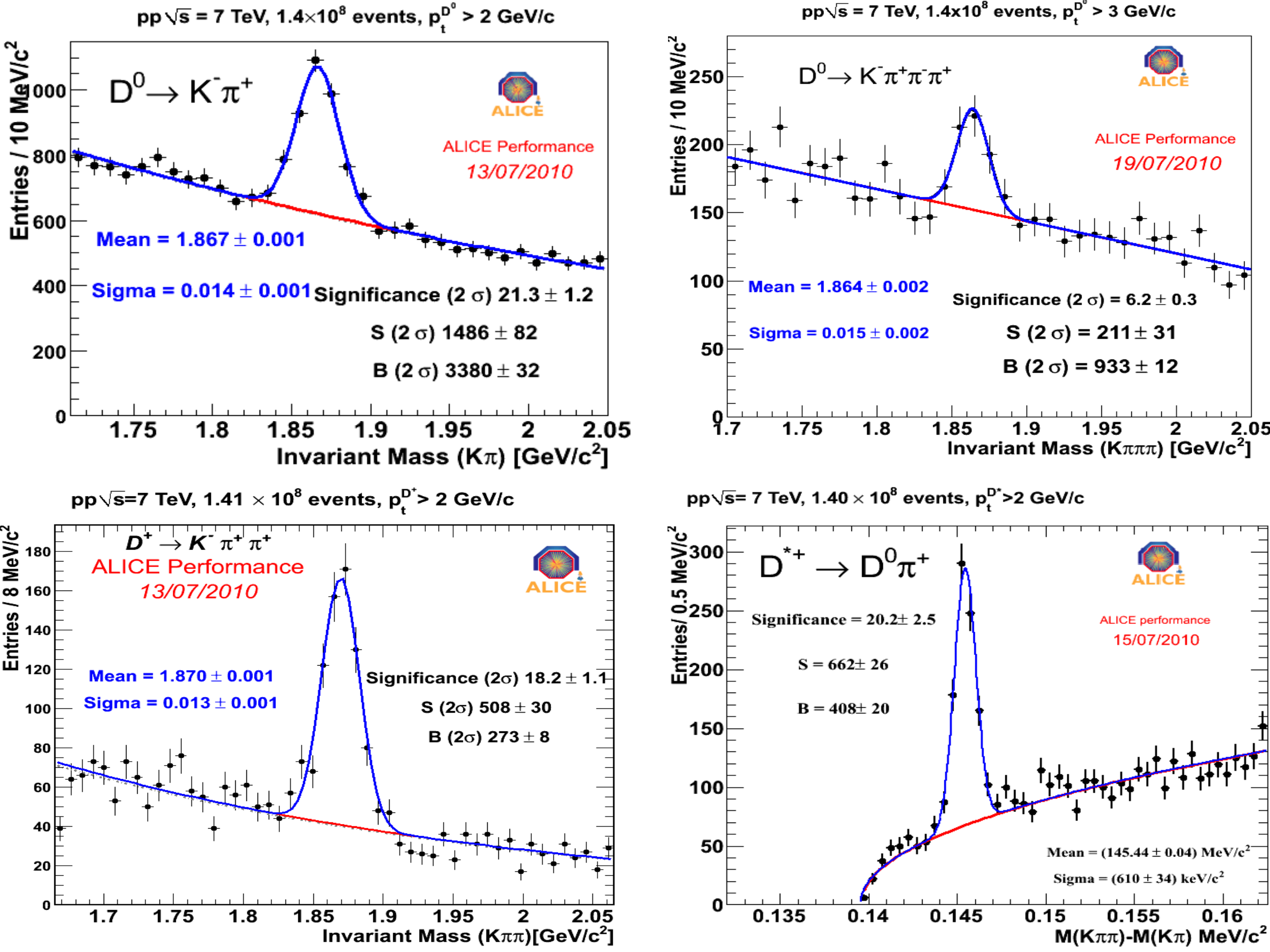}}

\caption{Invariant mass distributions for some exclusive hardonic charm decays.}
\label{f:charm}
\end{figure}

At the time of this conference, a number of physics results at both 0.9 TeV (2009 data) and 7 TeV have already been finalized and published and are shortly summarized in these proceedings, followed by some preliminary analysis results.

The charged particle density ${\rm d}N_{\rm ch}/{\rm d}\eta$ as well as the multiplicity distributions were measured at 0.9, 2.36 and 7 TeV for both inelastic as well as non-single diffractive collisions~\cite{:2009dt,Aamodt:2010pp,Aamodt:2010ft}. The energy dependence of the multiplicity is well described by a power law in energy, $s^{0.1}$, and increases significantly stronger than predicted by most event generators. Most of this stronger increase happens in the tail of the multiplicity distribution, i.e. for events with much larger than average multiplicity. Likewise, neither the transverse momentum distribution  at 900 GeV nor the dependence of average $p_T$ on $N_{\rm ch}$ is well described by various versions of generators~\cite{Aamodt:2010my}, in particular when including low momentum particles ($p_T < 500$ MeV). The shape of the $p_T$ spectrum as function of multiplicity hardly changes below  0.8 GeV (which includes the large majority of all particles), whereas the power law tail increases rapidly above about 1-2 GeV.

Bose Einstein (or Hanbury-Brown Twiss, HBT) correlations between identical particles can be used to measure the space-time evolution of the dense matter system created in heavy ion collisions; their interpretation in elementary reactions ($e^+e^-, pp$) is more controversial. However, their measurement at LHC energies is important also for pp as a comparison for the future heavy ion data, both because non-HBT correlations, which have to be subtracted, are expected to increase with energy (e.g. via increased jet and mini-jet activity in MB events) and in order clarify if systematic trends, seen e.g. as function of multiplicity and pair momentum, differ between pp and heavy ions. The trends observed at 900 GeV~\cite{Aamodt:2010jj} can be summarised as follows:
HBT radii, which are typically of order 1 fm in elementary collisions, increase smoothly with multiplicity (some 30\% over the $N_{\rm ch}$ range covered in our data) as previously observed both at lower (ISR, RHIC) and at higher (FNAL) energies,  a trend which is qualitatively similar to the one seen in heavy ion collisions. A quantitative comparison between pp and heavy ions is currently under way with the higher statistics data at 7 TeV. However, the decrease of radius with pair momentum seems to be much weaker in our data than the one seen at the Tevatron~\cite{E735}.  Already at 900 GeV we see, both in the data (with unlike-sign pairs) as well as in the Monte Carlo (both Phythia and Phojet), strong particle correlations at small relative pair momentum, i.e. in the region of the HBT effect, which had to be carefully studied and subtracted. If we evaluate the HBT radii without subtracting non-HBT correlations, as has been done traditionally in the past, we also do observe a strong momentum dependent decrease of the radius, as these minijet-like correlations become more pronounced a higher momentum, effectively simulating a decreasing radius.

At the LHC, by far the highest energy proton--proton collider, we have studied baryon transport over very large rapidity intervals 
by measuring the antiproton-to-proton ratio at mid-rapidity~\cite{Aamodt:2010dx} in order to discriminate between various theoretical models of baryon stopping.
Baryon number transport over large gaps in rapidity ($\Delta y = 8.92$ at 7 TeV) is often described in terms of a nonlinear three gluon configuration called 'baryon string junction'. The dependence of this process on the size of the rapidity gap has been a longstanding issue (for large gaps, where it should be dominant), with advocates for both very weak and rather strong dependencies.
In either case, the $\bar{p}/p$ ratio at LHC will be close to 1.0, with the difference between various models only of the order of a few percent. So this ratio has to be measured with high precision. While in the ratio many instrumental effects cancel, the very large difference between $p$ and $\bar{p}$ cross section for both elastic (track can get lost) and inelastic (particle can be absorbed)  reactions with the detector material lead to corrections of order 10\%, even in the very thin central part of ALICE. Therefore a very precise knowledge of the detector material as well as of the relevant cross section values at low momentum is required. The former was measured with the data via photon conversions with a relative precision of $< 7\%$ (absolute precision of better than 7 per mill radiation length!); the latter had to be cross checked with experimental data and the Fluka transport code as it turned out that all available versions of Geant overestimate the $\bar{p}$ cross sections by up to a factor of five. The $\bar{p}/p$ ratio was found to be compatible with 1.0 at 7 TeV and 4\% below 1.0 at 900 GeV, with an experimental uncertainty of about 1.5\%, dominated by the systematic error. The results favour models which predict a strong suppression of baryon transport over large gaps; they agree very well with standard event generators but not with those who have implemented enhanced proton stopping.

\begin{figure} {!ht}
 \centerline{\includegraphics[width=0.95\textwidth]{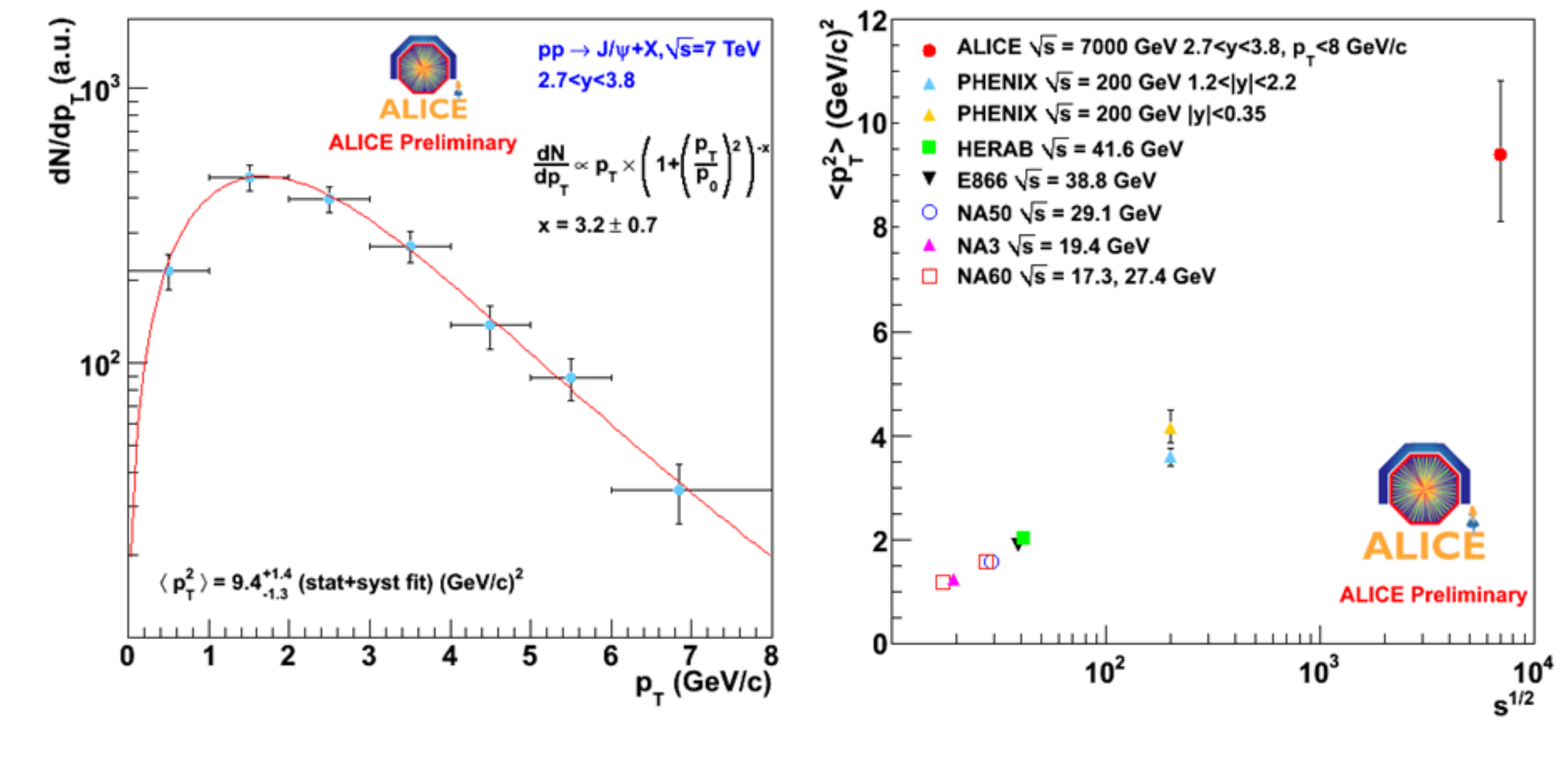}}

\caption{Left panel: $\Jpsi$ transverse momentum distribution measured in $2.7< y < 3.8$. Right panel: Average squared transverse momentum $<p_T^2>$ of $\Jpsi$ versus center of mass energy for the ALICE results at 7 TeV, compared to data at lower energies.}
\label{f:jpsi}
\end{figure}

Other physics topics which are currently being investigated in our large data sample of MB events include both soft (identified particle spectra and ratios, HBT) as well as semi-hard processes (heavy flavour semileptonic and hadronic decays, $\Jpsi$, particle correlations, jet fragmentation, event topology, underlying event studies, ..); they are described in more detail elsewhere in these proceedings~\cite{aliceparallel} and only shortly summarised in the following. 

The yield of strange particles 
($\Kzs,\Lambda, \Xi, \phi$) is shown in Figure~\ref{f:strange} as function of $p_T$ for the small 900 GeV data set taken in 2009. In most cases Phojet as well as several Phythia tunes are well below the data -- by factors of two to almost five -- and more so at high $p_T$ and for the heavier particles ($\Lambda, \Xi$). The ratio of $\Lambda$  to $\Kzs$, shown in Figure~\ref{f:klambda}, agrees very well with the STAR data at 200 GeV~\cite{Abelev:2006cs} but is significantly below the ratio measured by UA1~\cite{Bocquet:1995jq} and CDF~\cite{Abe:1989hy}. This discrepancy merits further investigation; it could be due to differences between the experiments in the acceptance, triggers, or correction for feed-down from weak decays. Baryon to meson ratios are of particular interest, as they rise well above 1 in nuclear collisions at RHIC. This 'meson-baryon' anomaly has been interpreted in the context of coalescence models as an indirect sign for the QGP, in which case it would not be obvious why similar ratios should be reached already in minimum bias pp interactions.

The measurement of hadronic charm cross sections is needed as baseline data for our nuclear program and of interest for comparison with pQCD because ALICE can measure charm production at midrapidity down to very low momenta. Figure~\ref{f:charm} shows the invariant mass distribution for a number of exclusive charm decays. The very good signal to background ratio for momenta as low as $p_T >$ 2 GeV is a consequence of both the use of PID information and the excellent performance of the ITS vertex detector, which with the current state of alignment has reached an impact parameter resolution of around $80\mim$ at $p_T$ = 1 GeV.  

$\Jpsi$ production has been measured in the muon spectrometer in the rapidity region $2.7 < y < 3.8$. In Fig.~\ref{f:jpsi} the $p_T$ distribution is shown on the left and the average $<p_T^2>$  as function of the center of mass energy on the right. The spectral shape is fully consistent with the preliminary data shown at this conference by LHCb in the same rapidity acceptance, and the average squared $p_T$ is seen to follow smoothly the rising trend known from lower energies.

\section{Summary}
After two decades of design, R\&D, construction, installation, commissioning and simulations, the ALICE experiment has 'hit the ground running' since LHC started its operation at the end of 2009. Most systems are fast approaching design performance, and physics analysis is well under way. While heavy ion physics will be its main subject, the collaboration has started to explore the 'terra incognita' at LHC with pp collisions, along the way gaining experience and sharpening its tools in anticipation of the first heavy ion run later this year.


\begin{thebibliography}{99}
%\cite{Carminati:2004fp}
\bibitem{Carminati:2004fp}
  F.~Carminati {\it et al.}  [ALICE Collaboration],
  %``ALICE: Physics performance report, volume I,''
  J.\ Phys.\ G {\bf 30} (2004) 1517.
  %%CITATION = JPHGB,G30,1517;%%

%\cite{Alessandro:2006yt}
\bibitem{Alessandro:2006yt}
  B.~Alessandro {\it et al.}  [ALICE Collaboration],
  %``ALICE: Physics performance report, volume II,''
  J.\ Phys.\ G {\bf 32} (2006) 1295.
  %%CITATION = JPHGB,G32,1295;%%
  
 %\cite{JSPLHC2010}
\bibitem{JSPLHC2010} 
J.~Schukraft in {\em  Proceedings of ``Physics at the LHC 2010''
}, 7-12 June 2010, DESY, Hamburg, Germany.
 
%\cite{Evans:2009zz}
\bibitem{Evans:2009zz}
  J. Schukraft and C. Fabjan in L.~Evans (Editor),
``The Large Hadron Collider: A marvel technology,''
%\href{http://www.slac.stanford.edu/spires/find/hep/www?irn=8688672}{SPIRES 
%entry}
{\it  Lausanne, Switzerland: EPFL (2009) 251 p}.
  

\bibitem{ALICEdet} 
K.~Aamodt {\it et al.}[ALICE Collaboration], 
J. Instrum \textbf{3}, S08002 (2008).

%\cite{Alme:2010ke}
\bibitem{Alme:2010ke}
  J.~Alme {\it et al.},
  %``The ALICE TPC, a large 3-dimensional tracking device with fast readout for
  %ultra-high multiplicity events,''
  arXiv:1001.1950 [physics.ins-det].

 
%\cite{:2009dt}
\bibitem{:2009dt}
  K.~Aamodt {\it et al.}  [ALICE Collaboration],
  %``First proton--proton collisions at the LHC as observed with the ALICE
  %detector: measurement of the charged particle pseudorapidity density at
  %sqrt(s) = 900 GeV,''
  Eur.\ Phys.\ J.\  C {\bf 65} (2010) 111.


%\cite{Aamodt:2010pp}
\bibitem{Aamodt:2010pp}
  K.~Aamodt {\it et al.}  [ALICE Collaboration],
  %``Charged-particle multiplicity measurement in proton-proton collisions at
  %sqrt(s) = 7 TeV with ALICE at LHC,''
  Eur.\ Phys.\ J.\  C {\bf 68} (2010) 345.
 
  
%\cite{Aamodt:2010ft}
\bibitem{Aamodt:2010ft}
  K.~Aamodt {\it et al.}  [ALICE Collaboration],
  %``Charged-particle multiplicity measurement in proton-proton collisions at
  %sqrt(s) = 0.9 and 2.36 TeV with ALICE at LHC,''
  Eur.\ Phys.\ J.\  C {\bf 68} (2010) 89.

  
%\cite{Aamodt:2010my}
\bibitem{Aamodt:2010my}
  K.~Aamodt {\it et al.}  [ALICE Collaboration],
  %``Transverse momentum spectra of charged particles in proton-proton
  %collisions at $\sqrt{s} = 900$~GeV with ALICE at the LHC,''
  Phys.\ Lett.\  B {\bf 693} (2010) 53.


%\cite{Aamodt:2010jj}
\bibitem{Aamodt:2010jj}
  K.~Aamodt {\it et al.}  [ALICE Collaboration],
  %``Two-pion Bose-Einstein correlations in pp collisions at sqrt(s)=900 GeV,''
Phys.\ Rev.\  D {\bf 82} (2010) 052001.
  %arXiv:1007.0516 [hep-th].

 
%\cite{E735}
\bibitem{E735}
  T.~Alexopoulos {\it et al.}  [E735 Collaboration],
  Phys.\ Rev.\  D {\bf 482} (1993) 1931.

%\cite{Aamodt:2010dx}
\bibitem{Aamodt:2010dx}
  A.~K.~Aamodt {\it et al.}  [ALICE Collaboration],
  %``Midrapidity antiproton-to-proton ratio in pp collisions at $\sqrt{s} = 0.9$
  %and $7$~TeV measured by the ALICE experiment,''
  Phys.\ Rev.\ Lett.\  {\bf 105} (2010) 072002.
  %[arXiv:1006.5432 [hep-ex]].


\bibitem{aliceparallel}
 see contributions by M. Lopez Noriega, A. Grelli, R. Bailhache, J. Castillo, G. Bruni, I. Belikov, and J. Rak in these proceedings.

%\cite{Abelev:2006cs}
\bibitem{Abelev:2006cs}
  B.~I.~Abelev {\it et al.}  [STAR Collaboration],
  %``Strange particle production in p + p collisions at s**(1/2) = 200-GeV,''
  Phys.\ Rev.\  C {\bf 75} (2007) 064901.

  
%\cite{Bocquet:1995jq}
\bibitem{Bocquet:1995jq}
  G.~Bocquet {\it et al.} [UA1 Collaboration],
  %``Inclusive production of strange particles $p \bar{p}$ collisions at
  %$\sqrt{s}$ = 630-GeV with UA1,''
  Phys.\ Lett.\  B {\bf 366} (1996) 441.

  
%\cite{Abe:1989hy}
\bibitem{Abe:1989hy}
  F.~Abe {\it et al.}  [CDF Collaboration],
  %``$K_s^0$ production in $\bar{p}p$ interactions at $\sqrt{s}=$ 630 GeV and
  %1800 GeV,''
  Phys.\ Rev.\  D {\bf 40} (1989) 3791; 
   D.~E.~Acosta {\it et al.}  [CDF Collaboration],
 %``$K^0_S$ and $\Lambda^0$ production studies in $p\bar{p}$ collisions at
  %$\sqrt{s}=$ 1800-GeV and 630-GeV,''
     Phys.\ Rev.\  D {\bf 72} (2005) 052001.
  

\end{thebibliography}
\end{document}